\documentclass{article}
\usepackage{PRIMEarxiv}

\usepackage[utf8]{inputenc} 
\usepackage{hyperref}       
\usepackage{url}            
\usepackage{booktabs}       
\usepackage{amsfonts}       
\usepackage{nicefrac}       
\usepackage{microtype}      
\usepackage{lipsum}
\usepackage{fancyhdr}       
\usepackage{graphicx}       
\graphicspath{{media/}}     

\usepackage{graphicx}
\usepackage[most]{tcolorbox}
\usepackage{caption}
\usepackage{subcaption}
\usepackage{longtable}

\usepackage{amsmath,amssymb,amsfonts}

\usepackage{textcomp}

\usepackage{verbatim}

\usepackage{xcolor}
\usepackage{url}
\usepackage{xr-hyper}
\usepackage{hyperref}
\usepackage{pdfpages} 
\usepackage{bm}
\usepackage{multirow}
\usepackage{multicol}
\usepackage{array}
\newcolumntype{L}{>{\arraybackslash}m{3cm}}

\usepackage{fullpage}
\usepackage{rotating}
\usepackage{stmaryrd}
\usepackage{proof}

\usepackage{tikz}
\usetikzlibrary{bayesnet}
\usetikzlibrary{decorations.pathmorphing} 
\usetikzlibrary{fit}
\usetikzlibrary{backgrounds}	

\usepackage{amsthm}

\usepackage[ruled]{algorithm2e}
\usepackage{algorithmic} 

\usepackage[style=numeric,sorting=none,firstinits=true]{biblatex}
\title{Feature Engineering Methods on Multivariate Time-Series Data for Financial Data Science Competitions
\thanks{\textit{\underline{Citation}}: 
\textbf{Authors. Title. Pages.... DOI:000000/11111.}} 
}

\author{
  Thomas Wong \\
  Imperial College London \\
  London\\
  \texttt{mw4315@ic.ac.uk} \\
   \And
  Mauricio Barahona \\
  Imperial College London \\
  London\\
  \texttt{m.barahona@imperial.ac.uk} \\
}

\begin{document}
\maketitle

\begin{abstract}
We apply different feature engineering methods for time-series to US market price data. The predictive power of models are tested against Numerai-Signals targets. 
\end{abstract}


\section{Introduction} 

Financial data are often available in the form of time series. These time series are often highly dimensional with complex relationships between them. 

The complexity of financial data can be demonstrated in different aspects. Firstly, training data are often limited and the number of features that researchers can create is often much greater than the number of observations. In some research, such as \cite{kelly2022virtue}, the ratio of the number of features over the number of observations, defined as model complexity can increase up to hundreds for financial instruments with a limited amount of history. Traditional setups in machine learning are not well-equipped for these data-scarce environments. Secondly, multicollinearity is very common in financial data and choosing suitable regularisation methods is a key part of model training. Thirdly, distribution shifts in data also hinder learning robust model parameters over time. It is well-known in finance that regime shifts can invalidate trading strategies. Therefore any robust machine learning models for financial forecasts require ways to deal with the non-stationarity of data. Moreover, correlation structure between features is often hard to estimate. For example, estimating the correlation structure of a basket of assets is non-trivial as the number of assets can easily exceed the length of price history. Dimensionality reduction methods are often required to simplify the problem. 

Researchers do not have a consensus on the best approach to handle the complexity of financial time series. The classical view on the bias-variance trade-off suggests using simple models to avoid over-fitting, especially for environments with a low signal-to-noise ratio. However, recent research \cite{kelly2022virtue} suggests using complex models through extensive feature engineering and model ensembling to take advantage of the "double descent" phenomenon of the curve of test loss with respect to the number of model parameters (as a measure of model complexity) in deep learning \cite{NakkiranPreetum2021Dddw}.

Financial time series can be treated directly using classic methods such as ARIMA models \cite{percival_walden_2020} and more recently through deep learning methods such as Temporal Fusion Transformers \cite{Bryan19}. However, such deep learning methods are easily over-fitted and lead to expensive retraining for financial data, which are inherently affected by regime changes and high stochasticity. 

Alternatively, one can use various feature engineering methods to transform these time series into \emph{tabular form} through a process sometimes called `de-trending' in the financial industry, where the characteristics of a financial asset at a particular time point, including features from its history, are represented by a single dimensional data row (i.e., a vector). In this representation, the time dimension is not considered explicitly, as the state of the system is captured through transformed features at each time point and the continuity of the temporal dimension is not used. 

For example, we can summarise the time series of the return of a stock with the mean and standard deviation over different look-back periods. Grouping these data rows for different financial assets into a table at a given time point we obtain a \emph{tabular dataset}. If the features are informative, this representation can be used for prediction tasks at each time point, and allow us to employ robust and widely tested ML algorithms that are applicable to tabular data.  


This paper is work done in parallel to another work performed by the same authors on the Numerai Classic tournament \cite{}. There are two major differences between the two papers, with different datasets and methodologies presented. 
Firstly, the two papers used different temporal tabular datasets. In this paper, we created our own dataset, with open-source documentation detailing how the tabular features are created from raw financial datasets. In the other paper, we used the dataset provided by Numerai for model training, the feature creation process is proprietary and it is impossible for outside researchers to recreate the dataset. Secondly, the two papers focused on different tasks in a machine-learning pipeline. In this paper, our focus is on dataset creation and feature engineering. In the other paper, the focus is on training robust machine learning models given standardised data and ways to post-process and combine predictions from well-known algorithms, namely feature neutralisation and model selection. 

The paper is organised as follows.  Section ~\ref{section:background} introduces the Numerai-Signals tournament. Section ~\ref{section:feature_eng} describes and discusses different feature engineering methods that can be applied on multivariate time-series. Section~\ref{section:data} describes the pipeline used to create features from various raw financial databases. Section~\ref{section:ML-models} describes the machine learning model training pipeline and performances of models trained from features created from various feature engineering methods.

\section{Numerai-Signals tournament} 
\label{section:background}

Numerai \cite{numerai} is a hedge fund that uses crowd-sourced models to trade a market-neutral global equities portfolio. Numerai-Signals \cite{numerai-signals} is a competition organised by the Numerai which requires data scientists need to bring their own data to create trading signals for stocks traded in the global market. Numerai-Signals simplifies the complicated trading process into a stock ranking problem. Each week, the user predictions are evaluated with Spearman's correlation with the actual stock ranking. Unlike Kaggle data science competitions which are evaluated at a fixed period, submissions are evaluated with live data in an ongoing process.

\paragraph{Prediction task:} The tournament task is to predict the \emph{stock rankings} each week, ordered from lowest to highest expected return. The scoring is based on Spearman's rank correlation 
of the predicted rankings with the main target label (`target-20d'). Hence there is a single overall score each week regardless of the number of stocks to predict each week. Participants are not scored on the accuracy of the ranking of each stock individually. 
Numerai uses the predicted rankings to construct a market-neutral portfolio which is traded every week (As of Dec 2022), i.e., the hedge fund buys and short-sells the same dollar amount of stocks. Therefore the relative return of stocks is more relevant than the absolute return, hence the prediction task is a ranking problem instead of a forecast problem.

\paragraph{Stock Return Targets:} Numerai provides five normalised targets (As of Dec 2022), which represent forward stock returns normalised against different factors at different time horizons. The tournament is scored against `target-20d', which represents 20 trading days normalised return against around 100 proprietary factors that include market, sector and Fama-French factors. 

\paragraph{Trading Universe:} An important feature of the stock universe provided by Numerai is that they are constructed in a point-in-time fashion, without the look-ahead bias which is common in academic financial research. The stock universe is constructed survivor bias free since 2003 and is updated on Friday every week. It takes into account real trading constraints such as liquidity and borrowing costs. The trading universe and targets provided by Numerai are very robust as they have incorporated risk management and operational considerations which are not addressed in academic research usually.

\section{Extracting features from multi-variate time series}
\label{section:feature_eng}

In this section, we introduce different mathematical transformations that are used to extract features from multi-variate time series such as price data. 

\paragraph{Multi-variate time-series}
A multi-variate time-series $X$ of $T$ steps and $N$ channels can be represented as $X = (\bm{x}_1, \bm{x}_2, \dots, \bm{x}_i, \dots, \bm{x}_T) $, with $1 \leq i \leq T$ and each vector $\bm{x}_i \in \mathbb{R}^N$ represents the values of the $N$ channels at time $i$. The number of channels of the time series is assumed to be fixed throughout time with regular and synchronous sampling, i.e. the values in each vector from multiple channels arrive at the same time at a fixed frequency. 

\paragraph{Feature extraction}
Feature extraction methods are defined as functions that map the two-dimensional time-series $X \in \mathbb{R}^{T \times N}$ to a one-dimensional feature space $f(X) \in \mathbb{R}^K$ where $K$ is the number of features. Feature extraction methods reduce the dimension and noise in time-series data. With feature extraction methods, traditional machine learning models such as gradient boosting decision trees can be used without relying on advanced neural network architectures such as Recurrent Neural Networks (RNN) or Long-Short-Term-Memory (LSTM) Networks. Computational resources can be reduced as simpler machine-learning models with fewer parameters can be used. 

\paragraph{Look-back windows}
For financial time series which can potentially grow with infinite size, a look-back window is used to restrict the data size when calculating features from time series. To avoid look-ahead bias, features that represent the state of time-series at time $i$ can only be calculated using values obtained up to time $i$, which is $(\bm{x}_1, \bm{x}_2, \dots, \bm{x}_i)$. In financial analysis, data collected more recently often have more importance than data collected from a more distant past. Therefore, analysis is often restricted to use the most recent $k$ data points only, which are $(\bm{x}_{i-k}, \bm{x}_{i+1-k}, \dots, \bm{x}_i)$. This represents the state of the financial time series at time $i$ with a look-back window of size $k$. Feature extraction methods are applied on data within the look-back window only. In practice, multiple look-back windows are used to extract features corresponding to short-term and long-term price trends. At each time $i$, features extracted with different look-back windows are concatenated to represent the state of the time series.

\subsection{Basic Statistics} 

Different statistical moments can be used to summarise data. Mean, variance, skewness and kurtosis which correspond to the first four statistical moments are widely used in machine learning applications. Moments higher than the fourth order are rarely used due to the lack of interpretability. 

At each time $i$ and a look-back window of size $k$, the statistical moments can be calculated on each channel of data within the look-back window $(\bm{x}_{i-k}, \bm{x}_{i+1-k}, \dots, \bm{x}_i)$. In this paper, moments up to the fourth order are calculated. For the multivariate time series with $N$ channels, statistical moments are calculated independently for each channel and a total of $4 \times N$ features are obtained at each time $i$.

\subsection{Catch22} 


Catch22 \cite{Lubba2019} is a general-purpose time-series feature extraction method. It is an optimised set of 22 features based on the 4791 features proposed in highly comparative time-series analysis (hctsa) \cite{hctsa}. Catch22 creates a set of diverse and interpretable features of time series in which all computations are deterministic. The computed features are scale-invariant, namely, it does not capture the location (mean) and spread (variance) properties of time series. Catch22 uses interdisciplinary methods to derive features of different themes, including data distribution, temporal statistics, linear and non-linear auto-correlations and entropy. Data distribution refers to statistical properties derived from histogram of numerical values in the time series. Temporal statistics refers to basic statistical properties of temporal trends in the time series such as the longest period of consecutive values above the mean. Linear and non-linear auto-correlations refer to different ways to measure auto-correlations in the time series with ideas from the Fourier spectrum and Auto Mutual Information. Entropy refers to Shannon entropy and other complexity measures. The description of how each of the features in Catch22 is calculated can be found in Table 1 in \cite{Lubba2019}. For a multivariate time series with $N$ channels, Catch22 is applied to each channel individually and generates a total of $22 \times N$ features at each time $i$.

\subsection{Signature Transforms} 

Signature transforms \cite{Lyons07, Chevyrev16, Terry22}, based on rough path theory, can be used to extract features from multi-variate time series. Signature transforms are applied on continuous paths. A path $X$ is defined as a continuous function from a finite interval $[a,b]$ to $\mathbb{R}^d$ with $d$ the dimension of the path. $X$ can be parameterised in coordinate form as $X_t = (X_t^1,X_t^2,\dots,X_t^d)$ with each $X_t^i$ being a single dimensional path. 

For each index $ 1 \leq i \leq d$, the increment of $i$-th coordinate of path at time $t \in [a,b]$, $S(X)_{a,t}^i$, is defined as 
\begin{equation*}
    S(X)_{a,t}^i = \int_{a<s<t} \mathrm{d}X_s^i = X_t^i - X_a^i
\end{equation*}
As $S(X)_{a,\cdot}^i$ is also a real-valued path, the integrals can be calculated iteratively. A $k$-fold iterated integral of $X$ along the indices $i_1,\dots,i_k$ is defined as 
\begin{equation*}
    S(X)_{a,t}^{i_1,\dots,i_k} = \int_{a<t_k<t} \dots \int_{a<t_1<t_2}   \mathrm{d}X_{t_1}^{i_1}  \dots \mathrm{d}X_{t_k}^{i_k} 
\end{equation*}

The Signature of a path $X: [a,b] \mapsto \mathbb{R}^d$, denoted by $S(X)_{a,b}$, is defined as the infinite series of all iterated integrals of $X$, which can be represented as follows 
\begin{align*}
    S(X)_{a,b} &= (1, S(X)_{a,b}^1, \dots, S(X)_{a,b}^d,  S(X)_{a,b}^{1,1}, \dots ) \\
                &=  \bigoplus_{n=1}^{\infty} S(X)_{a,b}^n
\end{align*}

An alternative definition of signature as the response of an exponential nonlinear system is given in \cite{Terry22}. 

Log Signature can be computed by taking the logarithm on the formal power series of Signature. No information is lost as it is possible to recover the (original) Signature from Log Signature by taking the exponential \cite{Chevyrev16,Terry22}. Log Signature provides a more compact representation of the time series than Signature. 
\begin{equation*}
    log S(X)_{a,b} =  \bigoplus_{n=1}^{\infty}  \frac{(-1)^{(n-1)}}{n} S(X)_{a,b}^{\bigotimes n} 
\end{equation*}

Signatures can be computed efficiently using the Python package signatory \cite{kidger2021signatory}. The signature is a multiplicative functional in which Chen's identity holds. This allows quick computation of signatures on overlapping slices in a path. Signatures provide a unique representation of a path which is invariant under reparameterisation \cite{Chevyrev16, Terry22}. Rough Path Theory suggests the signature of a path is a good candidate set of linear functionals which captures the aspects of the data necessary for forecasting. 



\section{Data Creation Pipeline}
\label{section:data}

\paragraph{Data Sources} 

Data from traditional sources, such as price and financials are used in addition to alternative datasets such as sentiment to create the feature set. Price data from CRSP is used to create different features using the above methods (stats, Catch22, signature). Financial data are sourced from Open Source Asset Pricing \cite{ChenZimmermann2021}. Sentiment data from Ravenpack are used. Financial Data are collected between 2003-01-31 and 2021-12-31.

\paragraph{Universe Creation} 

Numerai-Signals \cite{numerai} provides targets identified by Bloomberg tickers. Between 2003-01-31 and 2022-03-11, there are a total of 5842 US stock entries in the Numerai-Signals universe. 

A metadata table for our dataset is created by mapping the Bloomberg tickers with the key fields in various data sources. We map Bloomberg tickers to ticker names in CRSP, taking into account ticker name changes. We then use CUSIP and ISIN, which are two commonly used unique identifiers for US-traded stocks to map the stock entries to Compustat and Ravenpack databases. We obtain 5518 stock entries that can be validly mapped to any of the databases.

\paragraph{Targets}

We use `target-20d' provided by Numerai, which are 20 (trading) days forward return normalised against around 100 proprietary factors that include market, sector and Fama-French factors \cite{numerai-signals}. The target is scaled between 0 to 1, where 0 represents the quantile of the lowest return and 1 represents the quantile of the highest return.

\paragraph{Price Features}
Daily price data are used to calculate different features. Three different feature extraction approaches are used, including statistical, signature transforms \cite{kidger2021signatory} and Catch22 \cite{Lubba2019}. Pre-processing is applied to daily price data. The \textbf{average price} of a stock on each day is computed as the simple average of the dividend and splits adjusted open, high, low and close price. For each feature extraction approach, look-back windows with different lengths need to be defined to capture price patterns at different time resolutions. Look-back windows of sizes 21, 63 and 252 are chosen to capture short-term, mid-term and long-term price patterns in each feature extraction approach.

To calculate statistical features, the log-returns of the \textbf{average price} time series are first calculated, defined as
\begin{equation*}
    \text{log return}(x_t) = \frac{\log \text{average price}(x_t)}{\log\text{average price}(x_{t-1})}
\end{equation*}

On each trading day, the mean, standard deviation, skewness and kurtosis of log returns in different look-back windows are calculated. In total there are 12 statistical features.  The log of the \textbf{average price} is used to calculate the Catch22 features. In total there are 66 Catch22 features. Before applying signature transforms, the log of the \textbf{average price} is calculated to adjust for the effect of compounding growth in asset prices. Moving averages of the log \textbf{average price} with 5 and 21 days windows are added to the \textbf{average price} time series to create a lagged price time series, which is a multi-variate time-series with 3 channels. For each look-back window, the signature transforms up to the fourth level and is computed on the lagged price time series, which gives 32 log signatures. In total there are 96 log-signature features.

\paragraph{Sentiment Features}

Ravenpack collects and processes stock news in an easy-to-use format. For each piece of news, an `event relevance score' is given to indicate how relevant the news article is to the stock mentioned. On a scale between 0 to 100, we filter to include news that is the most relevant (an `event relevance score' of 100). We also further filter news to ensure uniqueness by including only news with an `event similar days' greater than 1, which means there is no similar news that occurred within a day. Each piece of news is given a sentiment score between -1 and 1, with negative scores indicating a negative outlook on the stock price and positive scores indicating a positive outlook on the stock price. 

For each stock, the \textbf{average sentiment} is calculated by the simple average of filtered news on each trading day. Moving averages of the \textbf{average sentiment} are also calculated with a look-back window of 21,63 and 252. This gives 4 features based on the overall sentiment. 

Each news article is also classified into different categories which correspond to different corporate events. Based on the data between 2003-01-03 and 2013-23-27 (the training and validation period for the machine learning pipeline described below), the top 200 categories of news that are most commonly found in stocks during the above period are selected for further study. Grouping the news by the top 200 categories, we then calculate the \textbf{average sentiment} using news in each selected category only on each trading day. On trading days that have no events of that category, the sentiment score is set to zero on that day. 252 days moving average of the sentiment scores by category are used instead of the actual time series due to the sparsity of events in each category. In total there are 204 sentiment features. 

\paragraph{Financial Features}
The monthly financial features obtained from Open Source Asset Pricing  \cite{ChenZimmermann2021} are re-sampled into weekly data. For example, financial features obtained at the end of Apr 2010 will be used in the 4 following weeks in May 2010. There are 204 financial features in their dataset, computed from different data sources such as company annual reports, analyst estimates and regulatory filings. Detailed implementation for financial features can be obtained from their paper and website \cite{ChenZimmermann2021}. 

\paragraph{Normalisation} 
Using the features calculated above, the features of each stock are normalised within each week. Rank transform is used to bin the features into 5 equal-sized quantiles. The transformed features are in the format of integers between -2 and 2, with -2 representing the 20\% of data with the lowest values in that feature and 2 representing the 20\% of data with the highest values in that feature. 

\section{Machine Learning Models}
\label{section:ML-models}

\paragraph{Data Split} 

We consider a walk-forward cross-validation approach to train different ML models with the latest data available. The training period uses an expanding window. The details are described in table \ref{table:cross-validation-online-expanding}. We apply a 1-year gap between the training and validation period to reduce the effect of recency bias so that the performance of the validation period will better reflect future performance. The gap between the validation period and the test period is set to 26 weeks to allow for sufficient time to deploy trained machine learning models.  

CV 0 is used for hyper-parameters optimisation and the optimised hyper-parameters are then used for the rest of the cross-validations (CV 1 to CV 4). The aim of this approach is to demonstrate the robustness of hyper-parameters and reduce computational costs to regularly update hyper-parameters. Hyper-parameters are robust if they can be transferred from previous cross-validations to later cross-validations without a significant drop in model performances.

\begin{table}[tbh]
\begin{tabular}{|l|l|l|l|l|l|l|}
\hline
     & Train Start & Train End  & Validation Start & Validation End & Test Start & Test End  \\ \hline
CV 0 & 2003-01-31  & 2010-12-31 & 2012-01-06       & 2015-12-25     & 2016-07-01 & 2021-12-31 \\ \hline     
CV 1 & 2003-01-31  & 2012-01-06 & 2013-01-11       & 2016-12-30     & 2017-06-30 & 2021-12-31 \\ \hline 
CV 2 & 2003-01-31  & 2013-01-04 & 2014-01-10       & 2017-12-29     & 2018-06-29 & 2021-12-31 \\ \hline 
CV 3 & 2003-01-31  & 2014-01-03 & 2015-01-09       & 2018-12-28     & 2019-06-28 & 2021-12-31 \\ \hline 
CV 4 & 2003-01-31  & 2015-01-02 & 2016-01-08       & 2019-12-27     & 2020-06-26 & 2021-12-31 \\ \hline 
\end{tabular}
\caption{Cross validation schemes to retrain machine learning models on different parts of the data based on expanding windows}
\label{table:cross-validation-online-expanding}
\end{table}

\paragraph{Model Training} 

Using different feature sets described as above, we perform hyper-parameter optimisation for each machine learning pipeline on the training and validation dataset. We use optuna \cite{Takuya19} to perform 100 iterations of the hyper-parameter search for the hyper-parameters of machine learning models based on the cross-validation data split of CV 0. The hyper-parameter space of the machine learning models are listed in Table \ref{numerai-signals-hyperspace-1}. 

The predicted stock rankings in each week are scored against the rankings of 20 trading days return `target-20d' using Spearman's correlation, which is defined as \textbf{Corr} in the tournament. A positive Spearman correlation represents a better alignment between predicted and actual rankings. We select the best parameters for each machine learning pipeline that have the highest Sharpe ratio of correlation with the `target-20d' in the validation period.  The Sharpe ratio is computed as the ratio of the average of \textbf{Corr} over the standard deviation of \textbf{Corr}. Using the best parameters obtained from CV 0, we train models using 20 different random seeds and report performances in the test period based on the average prediction for other cross-validations. Results for CV 1 are listed in Table \ref{table:numerai-signals-benchmark-test-cv1}. Performances in test periods from other cross-validations (CV 2 to CV 4) are listed in tables \ref{table:numerai-signals-benchmark-test-cv2},\ref{table:numerai-signals-benchmark-test-cv3},\ref{table:numerai-signals-benchmark-test-cv4} in the supplementary information.  

The Sharpe ratio and Calmar ratio are computed. The Sharpe ratio is computed as the ratio of Mean \textbf{Corr} over Volatility \textbf{Corr}. The Max Drawdown is defined as the maximum loss from the local peaks of the cumulative sum of \textbf{Corr}. The Calmar ratio is computed as the ratio of Mean \textbf{Corr} over Max Drawdown.

\subsection{Performances of models in different cross-validation}

In feature sets based on price data only (signature, Catch22, statistical), models trained with the `signature' feature set have the highest Sharpe ratio in all cross-validations (CV 1 to CV 4). Models trained with the `Catch22' feature set perform the worst in all cross-validations (CV 1 to CV 4) by having a lower Sharpe ratio and higher Max Drawdown. For price data that are highly non-stationary, Catch22 failed to capture the dynamic nature of financial time series and was over-fitted to temporal patterns. The Validation Sharpe ratio of Catch22 is 0.8026 which is the highest in all feature sets based on price data. The validation Sharpe ratio is 0.7215 and 0.4929 for `signature' and `statistical' respectively. It suggests Catch22 over-fits to specific patterns in the price data. 

Models trained with all three feature sets based on price data (signature+catch22+statistical) perform better than models trained with individual feature sets. Sharpe and Calmar ratios are higher in all cross-validations. Different feature extraction methods are complementary in nature which makes them good candidates for feature ensembling. The result is also consistent with the findings in \cite{kelly2022virtue} which suggests increasing model complexity improves predictive power of models.

Models trained with the `financials' feature set performed slightly better than models trained with price data only (signature, Catch22, statistical). As `target-20d' represents stock returns normalised against Fama-French factors \cite{}, such as Momentum and Value. Commonly used price and financial features are already taken into account, and therefore provide little value for predicting `target-20d'. 

Contrary to the common belief that more data/features improve model performances, models trained with only `sentiment' features perform significantly better than models trained with all 5 feature sets. Models trained with all feature sets can consider feature interactions but they are not useful in improving prediction for `target-20d'. This suggests to learn unique(orthogonal) trading signals, it is not necessary to include the known trading signals (such as Fama-French factors) in model training if the target is already normalised against those signals. ML models can be independently trained on each feature set and then combined. From the perspective of the organiser of the Numerai-Signals tournament, it demonstrates the feasibility of crowd-sourcing financial signals from the community as each contributor does not need to have access to data that others have to build a good signal. The burden on contributors to collect raw data and create signals can be greatly reduced as they are not strictly required to process traditional stock trading factors such as price and financials.

\begin{table}[tbh]
\begin{tabular}{|l|l|l|l|l|l|}
\hline
Feature Sets    & Sharpe &  Calmar & Mean   & Volatility & Max Drawdown \\ \hline
All                     & 0.5158 &  0.0736 & 0.0147 & 	0.0286 &	0.1996 \\ \hline
Signature+Catch22+Stats & 0.2595 &  0.0158 & 0.0085 & 	0.0327 &	0.5379 \\ \hline
Signature               & 0.1997 & 	0.0110 & 0.0061 & 	0.0307 &	0.5546 \\ \hline
Catch22                 & 0.1632 & 	0.0077 & 0.0048 & 	0.0294 &	0.6262 \\ \hline
Statistics              & 0.1600 & 	0.0132 & 0.0050 &	0.0310 &	0.3789 \\ \hline
Financials              & 0.3101 & 	0.0293 & 0.0082 &	0.0264 &	0.2794 \\ \hline
Sentiment               & 0.5324 & 	0.0816 & 0.0168 &	0.0315 &	0.2060 \\ \hline
\end{tabular}
\caption{Strategy Performance of LightGBM models trained on different feature sets for the Numerai-Signals tournament in the test period for CV 1. In addition to models trained with individual feature sets (stats, signature, Catch22, financials, sentiment), we also report the model trained with all 5 feature sets combined (all).  }
\label{table:numerai-signals-benchmark-test-cv1}
\end{table}


\newpage 
\section{Acknowledgments}
This was was supported in part by the Wellcome Trust under Grant 108908/B/15/Z and by the EPSRC under grant EP/N014529/1 funding the EPSRC Centre for Mathematics of Precision Healthcare at Imperial.

\newpage
\printbibliography

@article{kelly2022virtue,
  title={The Virtue of Complexity Everywhere},
  author={Kelly, Bryan T and Malamud, Semyon and Zhou, Kangying},
  journal={Available at SSRN},
  year={2022}
}

@article{NakkiranPreetum2021Dddw,
issn = {1742-5468},
journal = {Journal of statistical mechanics},
language = {eng},
number = {12},
pages = {124003-},
title = {Deep double descent: where bigger models and more data hurt},
volume = {2021},
year = {2021},
author = {Nakkiran, Preetum and Kaplun, Gal and Bansal, Yamini and Yang, Tristan and Barak, Boaz and Sutskever, Ilya},
}

@misc{Takuya19,
  doi = {10.48550/ARXIV.1907.10902},
  url = {https://arxiv.org/abs/1907.10902},
  author = {Akiba, Takuya and Sano, Shotaro and Yanase, Toshihiko and Ohta, Takeru and Koyama, Masanori},
  title = {Optuna: A Next-generation Hyperparameter Optimization Framework},
  publisher = {arXiv},
  year = {2019},
  copyright = {arXiv.org perpetual, non-exclusive license}
}

@Article{Lubba2019,
author={Lubba, Carl H.
and Sethi, Sarab S.
and Knaute, Philip
and Schultz, Simon R.
and Fulcher, Ben D.
and Jones, Nick S.},
title={catch22: CAnonical Time-series CHaracteristics},
journal={Data Mining and Knowledge Discovery},
year={2019},
month={Nov},
day={01},
volume={33},
number={6},
pages={1821-1852},
issn={1573-756X},
doi={10.1007/s10618-019-00647-x},
url={https://doi.org/10.1007/s10618-019-00647-x}
}

@article{hctsa,
publisher = {Royal Society, The},
title = {Highly comparative time-series analysis: the empirical structure of time series and their methods},
year = {2013-Apr},
author = {Fulcher, BD and Little, MA and Jones, NS},
language = {eng},
}

@misc{Chevyrev16,
  doi = {10.48550/ARXIV.1603.03788},
  url = {https://arxiv.org/abs/1603.03788},
  author = {Chevyrev, Ilya and Kormilitzin, Andrey},
  title = {A Primer on the Signature Method in Machine Learning},
  publisher = {arXiv},
  year = {2016},
  copyright = {arXiv.org perpetual, non-exclusive license}
}

@misc{Terry22,
  doi = {10.48550/ARXIV.2206.14674},
  url = {https://arxiv.org/abs/2206.14674},
  author = {Lyons, Terry and McLeod, Andrew D.},
  title = {Signature Methods in Machine Learning},
  publisher = {arXiv},
  year = {2022},
  copyright = {Creative Commons Attribution 4.0 International}
}

@inproceedings{
kidger2021signatory,
title={Signatory: differentiable computations of the signature and logsignature transforms, on both {\{}CPU{\}} and {\{}GPU{\}}},
author={Patrick Kidger and Terry Lyons},
booktitle={International Conference on Learning Representations},
year={2021},
url={https://openreview.net/forum?id=lqU2cs3Zca}
}

@inproceedings{Lyons07,
publisher = {Springer Berlin Heidelberg},
series = {École d'Été de Probabilités de Saint-Flour, 1908},
title = {Differential Equations Driven by Rough Paths : Ecole d’Eté de Probabilités de Saint-Flour XXXIV-2004 },
year = {2007},
author = {Lyons, Terry J.},
address = {Berlin, Heidelberg},
edition = {1st ed. 2007.},
isbn = {1-280-85347-6},
language = {eng},
}

@book{percival_walden_2020, 
place={Cambridge}, 
series={Cambridge Series in Statistical and Probabilistic Mathematics}, 
title={Spectral Analysis for Univariate Time Series}, 
%DOI={10.1017/9781139235723}, 
publisher={Cambridge University Press},
author={Percival, Donald B. and Walden, Andrew T.}, 
year={2020}, 
collection={Cambridge Series in Statistical and Probabilistic Mathematics},
}

@misc{Bryan19,
  doi = {10.48550/ARXIV.1912.09363},
  url = {https://arxiv.org/abs/1912.09363},
  author = {Lim, Bryan and Arik, Sercan O. and Loeff, Nicolas and Pfister, Tomas},
  title = {Temporal Fusion Transformers for Interpretable Multi-horizon Time Series Forecasting},
  publisher = {arXiv},
  year = {2019},
  copyright = {arXiv.org perpetual, non-exclusive license}
}

@article{ChenZimmermann2021,
  title={Open Source Cross Sectional Asset Pricing},
  author={Chen, Andrew Y. and Tom Zimmermann},
  journal={Critical Finance Review},
  year={Forthcoming}
}

@online{numerai,
  author="Numerai",
  title="{Numerai Hedge Fund}",
  url="https://numerai.fund/",
  note="(2022, Apr 12)",
}

@online{numerai-signals,
  author="Numerai",
  title="{Numerai Signals Competition}",
  url="https://signals.numer.ai/tournament",
  note="(2022, Apr 12)",
}

\section{Supplementary Information}

\subsection{Hyper-parameter search space for different ML models}
\label{section:optuna}

\begin{figure}[tbh]
\begin{itemize}
    \item Machine Learning 
    \begin{itemize}
        \item LightGBM-gbdt 
        \begin{itemize}
            \item n estimators: low:50, high:1000, step:50
            \item learning rate: low:0.005, high:0.1, log:True 
            \item min data in leaf: low:2500, high:40000, step:2500
            \item lambda l1: low:0.01, high: 1, log:True
            \item lambda l2: low:0.01, high: 1, log:True
            \item feature fraction: low:0.1, high:1, step:0.05
            \item bagging fraction: low:0.5, high:1, step:0.05
            \item bagging freq: low:10, high:50, step:10
        \end{itemize}     
    \end{itemize}        
\end{itemize}
\caption{Hyper-parameter Space for machine learning models }    
\label{numerai-signals-hyperspace-1}
\end{figure}

\subsection{Model Performances in different CV} 

\begin{table}[tbh]
\begin{tabular}{|l|l|l|l|l|l|}
\hline
Feature Sets    & Sharpe & Calmar & Mean & Volatility & Max Drawdown \\ \hline
All                     & 0.6706 & 0.1728 & 0.0187 & 0.0279 & 0.1082 \\ \hline
Signature+Catch22+Stats & 0.2778 & 0.0176 & 0.0086 & 0.0310 & 0.4876 \\ \hline
Signature               & 0.2191 & 0.0140 & 0.0062 & 0.0283 & 0.4440 \\ \hline
Catch22                 & 0.1470 & 0.0073 & 0.0043 & 0.0293 & 0.5911 \\ \hline
Statistics              & 0.1994 & 0.0135 & 0.0061 & 0.0308 & 0.4518 \\ \hline
Financials              & 0.2998 & 0.0206 & 0.0087 & 0.0290 & 0.4223 \\ \hline
Sentiment               & 0.7083 & 0.2329 & 0.0222 & 0.0313 & 0.0953 \\ \hline
\end{tabular}
\caption{Strategy Performance of LightGBM models trained on different feature sets for the Numerai-Signals tournament in the test period for CV 2. In addition to models trained with individual feature sets (stats,signature,Catch22,financials,sentiment), we also report the model trained with all the 5 feature sets combined (all).  }
\label{table:numerai-signals-benchmark-test-cv2}
\end{table}

\begin{table}[tbh]
\begin{tabular}{|l|l|l|l|l|l|}
\hline
Feature Sets    & Sharpe & Calmar & Mean & Volatility & Max Drawdown \\ \hline
All                     & 0.7886 & 0.3249 & 0.0218 & 0.0276 & 0.0671 \\ \hline
Signature+Catch22+Stats & 0.4544 & 0.0679 & 0.0140 & 0.0307 & 0.2062 \\ \hline
Signature               & 0.3672 & 0.0562 & 0.0105 & 0.0285 & 0.1868 \\ \hline
Catch22                 & 0.3522 & 0.0473 & 0.0099 & 0.0281 & 0.2093 \\ \hline
Statistics              & 0.3586 & 0.0377 & 0.0103 & 0.0287 & 0.2735 \\ \hline
Financials              & 0.4748 & 0.0723 & 0.0109 & 0.0229 & 0.1507 \\ \hline
Sentiment               & 0.7816 & 0.2505 & 0.0278 & 0.0356 & 0.1110 \\ \hline
\end{tabular}
\caption{Strategy Performance of LightGBM models trained on different feature sets for the Numerai-Signals tournament in the test period for CV 3. In addition to models trained with individual feature sets (stats,signature,Catch22,financials,sentiment), we also report the model trained with all the 5 feature sets combined (all).  }
\label{table:numerai-signals-benchmark-test-cv3}
\end{table}

\begin{table}[tbh]
\begin{tabular}{|l|l|l|l|l|l|}
\hline
Feature Sets    & Sharpe & Calmar & Mean & Volatility & Max Drawdown \\ \hline
All                     & 0.8342 & 0.3758 & 0.0239 & 0.0287 & 0.0636 \\ \hline
Signature+Catch22+Stats & 0.3763 & 0.0543 & 0.0123 & 0.0326 & 0.2265 \\ \hline
Signature               & 0.3561 & 0.0486 & 0.0104 & 0.0292 & 0.4116 \\ \hline
Catch22                 & 0.3473 & 0.0471 & 0.0101 & 0.0291 & 0.6742 \\ \hline
Statistics              & 0.2716 & 0.0279 & 0.0075 & 0.0276 & 0.4456 \\ \hline
Financials              & 0.3225 & 0.0468 & 0.0079 & 0.0245 & 0.1687 \\ \hline
Sentiment               & 0.9987 & 0.3846 & 0.0335 & 0.0335 & 0.0871 \\ \hline
\end{tabular}
\caption{Strategy Performance of LightGBM models trained on different feature sets for the Numerai-Signals tournament in the test period for CV 4. In addition to models trained with individual feature sets (stats,signature,Catch22,financials,sentiment), we also report the model trained with all the 5 feature sets combined (all).  }
\label{table:numerai-signals-benchmark-test-cv4}
\end{table}

\end{document}